\documentclass[a4paper,twocolumn,aps,superscriptaddress,floatfix]{revtex4}

\usepackage{amssymb,amsmath,stmaryrd}
\usepackage{bbm}
\usepackage{stix}
\usepackage{graphicx}
\usepackage{times}
\usepackage{float}
\usepackage{amsmath}
\usepackage{pst-node}
\usepackage{latexsym}
\usepackage{amsfonts}
\usepackage{amssymb}
\usepackage{makeidx}
\usepackage{cancel}
\usepackage{amsthm}
\usepackage{textgreek}
\usepackage[normalem]{ulem}
\usepackage[colorlinks,linkcolor=blue,citecolor=blue,urlcolor=blue]{hyperref}

\DeclareGraphicsExtensions{.pdf,.gif,.jpg}

\newcommand{\be}{\begin{equation}}
\newcommand{\ee}{\end{equation}}

\newcommand{\ba}{\[\begin{aligned}}
\newcommand{\ea}{\end{aligned}\]}

\newcommand{\bea}{\begin{eqnarray}}
\newcommand{\eea}{\end{eqnarray}}

\def\a{\alpha}
\def\b{\beta}

\def\d{\delta}

\def\e{\epsilon}

\def\l{\lambda}

\def\m{\mu}
\def\n{\nu}

\def\p{\pi}

\def\r{\rho}
\def\s{\sigma}

\def\t{\tau}

\def\F{\Phi}

\def\w{\omega}
\def\W{\Omega}
\def\q{\psi}
\def\Q{\Psi}

\def\blX{{\mathbf X}}
\def\blY{{\mathbf Y}}

\def\callL{\mbox{$\mathcal{L}$}}

\def\callN{\mbox{$\mathcal{N}$}}

\def\callY{\mbox{$\mathcal{Y}$}}

\def\iif{\infty}
\def\bra{\langle}
\def\ket{\rangle}

\def\Tr{{\rm Tr}}

\def\Im{{\rm Im}}

\def\1op{\hat{\mathbbm{1}}}
\def\nn{\nonumber}

\begin{document}
    
\title{From carriers and virtual excitons to 
exciton populations: Insights into
time-resolved ARPES spectra from an exactly solvable model}


\author{G. Stefanucci}
\affiliation{Dipartimento di Fisica, Universit\`{a} di Roma Tor Vergata,
Via della Ricerca Scientifica 1, 00133 Rome, Italy}
\affiliation{INFN, Sezione di Roma Tor Vergata, Via della Ricerca
  Scientifica 1, 00133 Rome, Italy}
  
\author{E. Perfetto}
\affiliation{Dipartimento di Fisica, Universit\`{a} di Roma Tor Vergata,
Via della Ricerca Scientifica 1, 00133 Rome, Italy}
\affiliation{INFN, Sezione di Roma Tor Vergata, Via della Ricerca
  Scientifica 1, 00133 Rome, Italy}


\begin{abstract}
We calculate the {\em exact} time-resolved ARPES spectrum of a 
two-band model semiconductor driven out of equilibrium by resonant 
and nonresonant laser pulses, highlighting the effects of 
phonon-induced decoherence and relaxation. 
{\em Resonant} excitations initially yield a replica of the {\em valence} band 
shifted upward by the energy of the exciton peak in photoabsorption. This  
phase is eventually destroyed by phonon-induced decoherence: 
the valence-band replica lowers in energy by the Stokes 
shift, locating at the energy of the exciton peak in 
photoluminescence, and its width grows  due to 
phonon dressing. {\em Nonresonant} excitations initially yield a map of the 
conduction band. Then electrons transfer their excess energy to the 
lattice and  bind with the holes left behind to form excitons. In this 
relaxed regime a replica of the {\em conduction} band appears inside 
the gap. At fixed momentum the lineshape of the conduction-band replica versus the 
photoelectron energy is proportional to 
the exciton wavefunction in ``energy space'' and it is 
highly asymmetric. Although the two-band 
model represents an oversimplified description of real materials the 
highlighted features are qualitative in nature; hence they provide useful 
insights into time-resolved ARPES spectra and their physical 
interpretation.

\end{abstract}

\maketitle

\section{Introduction}

Time-resolved and angle-resolved photoemission spectroscopy (trARPES) 
is currently one of the most flourishing playground for condensed 
matter theoreticians. Accurate approximations to the  
fundamental equations of many-body theory are being incessantly 
developed to relate the intensity and direction of the 
photocurrent to the behavior of quantum matter in equilibrium as well as  
nonequilibrium conditions.
The relationship between pseudogap, charge ordering and Fermi 
arcs in high temperature 
superconductors~\cite{Damascelli_RevModPhys.75.473,Vishik_2018}, 
Auger scattering, electron-phonon coupling, plasmonic excitations, and local screening 
in core excited 
metals~\cite{Doniach_1970,Citrin_PRB1977,Hufner-book}, carrier 
populations and conduction states in excited 
semiconductors~\cite{Rossnagel.2012,SangalliEPL2015,Smallwood2016,Mo2017,Caruso_PhysRevB.101.035128}, 
topological order~\cite{Lv2019}, excitonic insulator phases~\cite{Wakisaka_PhysRevLett.103.026402,Seki_PhysRevB.90.155116}, 
excitonic Mott transitions~\cite{Dendzik_PhysRevLett.125.096401} 
and exciton 
dynamics~\cite{Weinelt_PhysRevLett.92.126801,Suzuki_PhysRevLett.103.057401,ZHU201575,Varene_PhysRevLett.109.207601,Deinert_PhysRevLett.113.057602,PSMS.2016,Steinhoff2017,RustagiKemper2018,PSMS.2019,Christiansen_PhysRevB.100.205401} in pumped semiconductors is a 
non exhaustive list of the plethora of different phenomena leaving 
distinctive footprints in trARPES spectra.  

The interest in the exciton dynamics of (low-dimensional) 
semiconductors is steadily gaining momentum, especially due to 
potentially revolutionary applications in optoelectronics~\cite{Xiao2017}.
However,  the physical interpretation of trARPES spectra is still 
subject of debates. The main difficulty in developing accurate many-body schemes 
suitable for numerical implementations lies 
in the treatment of electron-electron and 
electron-phonon scatterings under nonequilibrium conditions. 
We find useful to distinguish between two different nonequilibrium regimes: the linear-response 
regime, where the photocurrent is proportional to the intensity of the 
{\em pump} field, and the nonlinear regime. Making predictions in the nonlinear 
regime is certainly more difficult. Calculations are often limited to the 
quasi-equilibrium state of matter, a condition which allows for 
introducing more or less controlled 
approximations like, e.g., the invariance under time translations and 
the fulfillment of the fluctuation-dissipation theorem in different 
bands. The linear regime 
offers a wider set of theoretical tools; 
exciton formation, coherence and relaxation can be addressed in a more rigorous  
framework.

The trARPES signatures of excitons in {\em linearly} excited semiconductors 
is the topic of this work. In fact,
there still are a few open questions pertaining with two distinct types of 
excitations. {\em Resonant} excitations are those of pump 
pulses with the 
same frequency as the energy $E^{(\rm x)}$ of a bright exciton. 
To avoid possible misinterpretations, 
by $E^{(\rm x)}$ we here denote the energy position of the excitonic 
peak in the photoabsorption spectrum of the ground-state system. Just after pumping a low-density gas of 
coherent excitons, also called nonequilibrium excitonic 
insulator or BEC exciton superfluid,
forms~\cite{Ostreich_1993,Hannewald-Bechstedt_2000,SzymaPRL2006,Hanai2016,HanaiPRB2017,Becker_PhysRevB.99.035304}. 
Theoretical works~\cite{Kremp_PhysRevB.78.125315,PSMS.2016,RustagiKemper2018,PSMS.2019,PBS.2020}
predict an excitonic sideband in the ARPES spectrum -- more 
precisely a replica of the {\em valence} band shifted upward by $E^{(\rm 
x)}$.
The subsequent dynamics of the coherent exciton gas involves 
electron-phonon scatterings~\cite{Hanai2016,Hannewald_PhysRevB.62.4519}, 
generation of phonons and phonon-induced decoherence~\cite{murakami2019ultrafast}, until the 
formation of a gas of incoherent excitons dressed by phonons, i.e.,
exciton-polarons.
This incoherent phase is expected to set in on a
time scale of a few hundreds of 
femtoseconds~\cite{Sundaram2002,Banyai_PhysRevLett.75.2188,BARAD199783,SangalliMariniJOPCS_2015} and it lasts until 
excitons radiatively recombine~\cite{Palummo-2015,WangChernikov2018}. 
To the best of our knowledge no theoretical 
ARPES studies exist in this phase. 
Is the excitonic sideband still visible? If so, is it any different 
from that of the coherent phase? Does the conduction band appear? 
Does the phonon bath relax? 
Besides strengthening our understanding
 an answer to these questions is becoming urgent.
Recent experiments have indeed reported excitonic sidebands in 
a resonantly pumped WSe$_{2}$ monolayer after 0.5 picoseconds~\cite{man2020experimental}
and in a WSe$_{2}$ bulk until 0.1 picoseconds~\cite{dong2020measurement}.

The second type of excitation is generated by {\em nonresonant} 
pumping.
Here the pump frequency is larger than the gap and 
electrons are promoted to empty conduction states.
Just after pumping the ARPES spectrum  provides a map of the 
conduction bands, the signal intensity being proportional to the 
band-resolved and momentum-resolved carrier populations. The excited electrons soon 
transfer their energy to the lattice, migrate toward the bottom of 
the conduction band~\cite{SangalliEPL2015,Caruso_PhysRevB.101.035128} and eventually bind with 
the left behind hole to 
form excitons~\cite{Weinelt_PhysRevLett.92.126801,Deinert_PhysRevLett.113.057602}. The spectral function in this relaxed (and incoherent) phase has been studied 
theoretically using 
the T-matrix approximation in the particle-hole 
channel assuming a quasi-thermal 
equilibrium~\cite{Kraeftbook,Schepe1998,Piermarocchi_PhysRevB.63.245308,Kremp_PhysRevB.78.125315,Kwong_PhysRevB.79.155205,Asano-2014,PSMS.2016,Steinhoff2017}. The 
theory predicts  an 
excitonic sideband inside the gap, about $E^{(\rm x)}$ above the 
valence band maximum. However, this spectral structure turns out to be a replica of the 
{\em conduction} band~\cite{Asano-2014,PSMS.2016,Steinhoff2017}. 
Could this be the ultimate fate of the excitonic 
sideband for resonant excitations? May different excitations 
(resonant versus nonresonant) yield different sidebands (valence 
replica versus conduction replica) in 
the relaxed and incoherent phase? Couldn't the replica of the conduction 
band be an artifact of the T-matrix approximation in combination with 
the assumption of quasi-thermal equilibrium?

We address the above issues through the {\em 
exact} analytic solution of a two-band model semiconductor where both 
electron-electron and electron-phonon interactions are taken into 
account. As the answer to the asked
questions is qualitative in nature, our results provide a
useful reference for benchmarking approximate many-body treatments.
The main findings of our investigation are: (i)~Incoherent excitons 
forming after resonant pumping do only change  slightly the replica of the 
valence band (observed in the coherent phase); in particular the 
replica lowers in 
energy by the Stokes shift~\cite{toyozawa_2003} and its width grows 
due to phonon dressing (exciton-polarons); (ii)~Phonon dressing is 
much faster than decoherence, the former being dictated by the 
largest phonon frequency whereas the latter by the smallest polaronic 
shift; (iii)~Excitons forming after nonresonant pumping give rise to a replica of 
the conduction band, thus confirming the results of previous 
studies~\cite{Asano-2014,PSMS.2016,Steinhoff2017}; 
for any fixed momentum the lineshape of the 
replica is determined by the 
exciton wavefunction in ``energy space'' and it is 
highly asymmetric.

The plan of the paper is as follows. In Section~\ref{epmodelsec} we 
introduce the model and set up the problem; we also discuss 
the behavior of relevant observable quantities in different scenarios. 
The exact solution of the model for both acoustic and optical phonons is derived in 
Section~\ref{exactsec}. We defer the reader to the Appendices for the 
calculation of the spectral function, momentum-resolved electron 
occupations, polarization and phonon occupations using the exact 
many-body wavefunction. Results for resonant and nonresonant excitations are 
presented in Sections ~\ref{resonantsec} and \ref{nonresonantsec} 
respectively. A summary of the main findings is drawn in 
Section~\ref{consec}.

\section{Exciton-polaron model}
\label{epmodelsec}

In standard notation the two-band model Hamiltonian reads
\begin{align}
\hat{H}&=\sum_{k}\big(\e^{c}_{k}\hat{c}^{\dag}_{k}\hat{c}_{k}+ 
\e^{v}_{k}\hat{v}^{\dag}_{k}\hat{v}_{k}\big)+\sum_{q}\w_{q}\hat{b}^{\dag}_{q}\hat{b}_{q}
\nn\\
&-\frac{1}{\callN}\sum_{pk_{1}k_{2}}U_{p}\hat{c}^{\dag}_{k_{1}+p}
\hat{v}_{k_{2}+p}
\hat{v}^{\dag}_{k_{2}}\hat{c}_{k_{1}}
\nn\\
&+\frac{1}{\sqrt{\callN}}\sum_{kq}
\l_{q}^{k}\big(\hat{c}^{\dag}_{k+q}\hat{c}_{k}\hat{b}_{q}+
\hat{c}^{\dag}_{k}\hat{c}_{k+q}\hat{b}^{\dag}_{q}\big).
\label{ham}
\end{align}
The first two terms describe free electrons in the valence ($v$) or 
conduction ($c$) band and free phonons. The remaining terms 
account for the electron-hole ($eh$) Coulomb interaction and the 
electron-phonon interaction; $\callN$ is the number of discretized 
momenta in the first Brillouin zone. In our model only conduction electrons 
interact with phonons. However, the idea presented 
below  can easily be adapted to include an interaction with 
valence electrons. 
The Hamiltonian in Eq.~(\ref{ham}) has been studied numerically 
with Quantum Monte Carlo 
methods to address the dependence of the exciton-polaron 
wavefunction on the electron-phonon 
coupling~\cite{Hohenadler_PhysRevB.76.184303,Burovski_PhysRevLett.101.116403}.
We are not aware of other exact numerical or analytical treatments.

The state $|\F_{0}\ket$ with a filled 
valence band, an empty conduction band and no phonons is an 
eigenstate of $\hat{H}$. We assume that the interaction $U_{p}$ is 
much larger than the energy gap $E_{g}$ between the conduction and 
valence bands. Then $|\F_{0}\ket$ is the ground state and, without any loss 
of generality, we set to zero its energy. 
We now consider an ultrafast and low-intensity laser pulse pumping electrons from the 
valence band to the conduction band. To lowest order in the light intensity 
the state of the system at 
the end of the pulse can be written as
\be
|\Q\ket=\a|\F_{0}\ket+\b|\F_{\rm x}\ket,
\label{cohstate}
\ee
with $\b=\sqrt{1-\a^{2}}\ll 1$ and 
\be
|\F_{\rm x}\ket=\sum_{k}Y^{0}_{k}\hat{c}^{\dag}_{k}\hat{v}_{k}|\F_{0}\ket
\label{fxt0}
\ee
the component of the full many-body state with one electron in the 
conduction band, one hole in the valence band and no phonons.
The coefficients
$Y^{0}_{k}$ depend on 
the laser pulse parameters, e.g., duration and frequency. The state 
in Eq.~(\ref{cohstate}) is not, in general, an eigenstate of $\hat{H}$. The pumped 
electron feels  the attractive interaction with the hole left 
behind and it is scattered by phonons. We shall investigate two 
different physical scenario.

In the so called resonant  case 
electrons are pumped at the exciton frequency and the state 
$|\F_{\rm x}\ket$ is a bright excitonic eigenstate of the electronic part 
of $\hat{H}$. Discarding the electron-phonon interaction and 
denoting by $E^{(\rm x)}< E_{g}$ the exciton energy the evolution of 
the state $|\Q\ket$ would simply be 
\be
|\Q(t)\ket=\a|\F_{0}\ket+\b 
e^{-iE^{(\rm x)}t}|\F_{\rm x}\ket.
\label{virtex}
\ee
As $|\Q(t)\ket$ has no phonons the electronic density matrix is a 
{\em pure state} 
\be
\hat{\r}_{\rm el}(t)\equiv \Tr_{\rm ph}\big\{\,|\Q(t)\ket\bra\Q(t)|\,\big\}
=|\Q(t)\ket\bra\Q(t)|,
\label{purestate}
\ee
where $\Tr_{\rm ph}$ signifies a trace over the phononic degrees of 
freedom. A system described by Eq.~(\ref{purestate}) is said to 
contain  {\em virtual} 
or {\em coherent} excitons~\cite{HaugKochbook,ShaferWegenerbook}. In fact, it is characterized by 
coherent oscillations of the polarization since the quantity
\be
\Tr_{\rm el}\left[\hat{\r}_{\rm el}(t)
\hat{v}^{\dag}_{k}\hat{c}_{k}\right]=\a^{\ast}\b \,
Y^{0}_{k}e^{-iE^{(\rm x)}t}
\ee
oscillates at the exciton frequency for all $k$'s, see also 
Eq.~(\ref{polariz}) below. In Ref.~\cite{PBS.2020} we argued that 
these coherent oscillations could be observed in trARPES using 
ultrafast probes of duration shorter than $2\p/E^{(\rm x)}$.

The state in 
Eq.~(\ref{virtex}) approximates the true time-dependent state 
only in the early stage of the evolution.
Just after pumping electrons and phonons begin to scatter, 
mutually dressing each others, and the initial coherence is eventually 
destroyed. The 
electronic system is expected to evolve toward an {\em admixture} of $|\F_{0}\ket$ and 
some exciton-like states $|\F^{(i)}_{\rm x}\ket$:
\begin{align}
\hat{\r}_{\rm el}(t\to\iif)&=\lim_{t\to\iif}\Tr_{\rm ph}\big\{\,|\Q(t)\ket\bra\Q(t)|\,\big\}
\nn\\ &=
|\a|^{2}\,|\F_{0}\ket\bra\F_{0}|
+|\b|^{2}\,\sum_{i}w_{i}|\F^{(i)}_{\rm x}\ket\bra\F^{(i)}_{\rm x}|.
\label{incoherentreg}
\end{align}
In this steady-state regime excitons are 
said {\em real} or {\em incoherent} and one can introduce the concept 
of {\em exciton populations} since the polarization does no longer 
oscillate. However, the existence and the  
characterization of the steady-state regime has so far been based on 
reasonable assumptions and it is still subject of debate. 
The purpose of this work is to provide useful 
insights into this issue through the exact solution of the 
time-dependent Schr\"odinger equation. 

In the second scenario 
the laser pulse generates free carriers in the conduction band
(nonresonant pumping). 
It is then expected that carriers give their excess energy to the 
lattice, thereby migrating toward the bottom of the conduction band and 
eventually binding with the holes left behind to form excitons. 
The phonon-driven formation of excitons is another debated 
topic in the literature as no real-time calculations are available to 
confirm this picture. Due to the simplicity of the model we could 
only address the dynamics of free carriers initially at the bottom of the 
conduction band.

Independently of the scenario we need to calculate the time-evolved state 
\begin{align}
|\Q(t)\ket&\equiv e^{-i\hat{H}t}|\Q\ket
=\a |\F_{0}\ket+\b  |\F_{\rm x}(t)\ket,
\label{tdPsi}
\end{align}
with
\be
|\F_{\rm x}(t)\ket=e^{-i\hat{H}t}\sum_{k}Y^{0}_{k}
\hat{c}^{\dag}_{k}\hat{v}_{k}|\F_{0}\ket.
\label{Phix}
\ee
Henceforth we refer to 
$|\F_{\rm x}(t)\ket$ as the exciton-polaron state although it may 
also describe unbound $eh$ pairs dressed by phonons.
From $|\F_{\rm x}(t)\ket$ we can 
monitor the momentum-resolved phonon occupations
\begin{align}
n_{q}(t)=\bra\F_{\rm x}(t)|\hat{b}^{\dag}_{q}\hat{b}_{q}
|\F_{\rm x}(t)\ket,
\label{nq(t)}
\end{align}
as well as the standard deviation of the phonon momentum 
\begin{align}
\s^{2}_{Q}(t)&=\bra\F_{\rm x}(t)|\hat{Q}^{2}|\F_{\rm 
x}(t)\ket-Q^{2}(t),
\end{align}
where $\hat{Q}=\sum_{q}q\,\hat{n}_{q}$ and $Q(t)=\sum_{q}q\,n_{q}$.
We can also calculate the Green's functions
\begin{align}
G^{cc}_{k}(t,t')&=i\bra\Q(t')|\hat{c}^{\dag}_{k}
e^{-i\hat{H}(t'-t)}\hat{c}_{k}|\Q(t)\ket
\nn\\
&=i|\b|^{2}\bra\F_{\rm x}(t')|
\hat{c}^{\dag}_{k}
e^{-i\hat{H}(t'-t)}\hat{c}_{k}|\F_{\rm x}(t)\ket
\label{Gcc}
\end{align}
and
\begin{align}
G^{cv}_{k}(t,t')&=i\bra\Q(t')|\hat{v}^{\dag}_{k}
e^{-i\hat{H}(t'-t)}\hat{c}_{k}|\Q(t)\ket,
\nn\\
&=i\a^{\ast}\b e^{iE_{0}t'}\bra\F_{0}|
\hat{v}^{\dag}_{k}
e^{-i\hat{H}(t'-t)}\hat{c}_{k}|\F_{\rm x}(t)\ket.
\label{Gcv}
\end{align}

The Green's functions in Eqs.~(\ref{Gcc}) and (\ref{Gcv})
contain information on the electronic 
transient 
spectrum, occupations and polarization.
In fact, the ARPES signal of the system at time $T$ is proportional 
to the transient spectral function $A_{k}(T,\w)$ which is in turn
related to the Green's function in 
Eq.~(\ref{Gcc}) through (for $\w$ above the valence band maximum) 
\be
A_{k}(T,\w)=-i\int d\t e^{i\w\t}G^{cc}_{k}(T+\frac{\t}{2},T-\frac{\t}{2}).
\label{spectralfunction}
\ee
The electron occupations in the conduction band can also be obtained 
from the same Green's function since
\be
n^{c}_{k}(t)=\bra\Q(t)|\hat{c}_{k}^{\dag}\hat{c}_{k}|\Q(t)\ket=-iG^{cc}_{k}(t,t).
\label{nc}
\ee
Denoting by $d_{k}$ the dipole matrix element between a 
conduction state and a valence state of momentum $k$ the polarization 
at time $t$ reads 
\begin{align}
P(t)&=\frac{1}{\callN}\sum_{k}d_{k}\bra\Q(t)|\hat{v}^{\dag}_{k}\hat{c}_{k}|\Q(t)\ket 
+{\rm h.c.}
\nn\\
&=-\frac{i}{\callN}\sum_{k}d_{k}G^{cv}_{k}(t,t)+{\rm h.c.}\;.
\label{polariz}
\end{align}
In the next section we expand the exciton-polaron state on a convenient basis and 
calculate the time-dependent expansion coefficients.

\section{Time-dependent exciton-polaron wave-function}
\label{exactsec}

We introduce the states
\be
|kq_{1}\ldots q_{M}\ket=
\hat{c}^{\dag}_{k-q_{1}\ldots-q_{M}}\hat{v}_{k}
\hat{b}^{\dag}_{q_{1}}\ldots\hat{b}^{\dag}_{q_{M}}|\F_{0}\ket,
\label{basis}
\ee
describing one $eh$ pair and $M$ phonons.
In terms of these states Eq.~(\ref{fxt0}) can be written as 
$|\F_{\rm x}\ket=\sum_{k}Y^{0}_{k}|k\ket$.
Hence the calculation of $|\F_{\rm x}(t)\ket$ passes through the calculation 
of $e^{-i\hat{H}t}|k\ket$. 
The Hamiltonian $\hat{H}$ maps the space 
spanned by the states  of Eq.~(\ref{basis}) onto the same space since
\begin{align}
\hat{H}|kq_{1}\ldots q_{M}\ket&=\big(
\e^{c}_{k-q_{1}\ldots-q_{M}}-\e^{v}_{k}+\sum_{j=1}^{M}\w_{q_{j}}\big)|kq_{1}\ldots q_{M}\ket
\nn\\
&-\frac{1}{\callN}\sum_{p}U_{p}|k+pq_{1}\ldots q_{M}\ket
\nn\\
&+\frac{1}{\sqrt{\callN}}\sum_{j=1}^{M}\l^{k}_{q_{j}}
|kq_{1}\ldots\stackrel{\sqcap}{q}_{j}\ldots q_{M}\ket
\nn\\
&+\frac{1}{\sqrt{\callN}}
\sum_{q}\l^{k}_{q}|kqq_{1}\ldots q_{M}\ket,
\label{Hkq1qM}
\end{align}
where the square-cap symbol ``$\stackrel{\sqcap}{}$'' signifies that the index 
below it is missing. We can therefore expand the exciton-polaron state as
\be
|\F_{\rm x}(t)\ket=\sum_{M=0}^{\iif}\frac{1}{M!}
\sum_{k,q_{1}\ldots q_{M}}Y_{kq_{1}\ldots q_{M}}(t)|kq_{1}\ldots q_{M}\ket.
\label{expansion}
\ee
Without any loss of generality we take the amplitudes 
$Y_{kq_{1}\ldots q_{M}}$ totally symmetric under a permutation of the 
phonon indices $\{q_{1},\ldots,q_{M}\}$ -- only this irreducible representation 
contributes in the sum of 
Eq.~(\ref{expansion}). Using the inner product,
\be
\bra k'q_{1}'\ldots q'_{M}|kq_{1}\ldots q_{M}\ket=
\d_{k'k}\sum_{P}\prod_{j=1}^{M}\d_{q'_{j}q_{P(j)}},
\label{innerproduct}
\ee
where the sum runs over all permutations of $\{1,\ldots,M\}$, we find
\be
\bra kq_{1}\ldots q_{M}| e^{-i\hat{H}t} |\F_{\rm x}\ket=
Y_{kq_{1}\ldots q_{M}}(t).
\label{amplitude}
\ee

Equations (\ref{Hkq1qM}) and (\ref{amplitude}) allows for generating 
a hierarchy of differential equations for the amplitudes
\begin{align}
i\dot{Y}_{kq_{1}\ldots q_{M}}(t)&=
\bra kq_{1}\ldots q_{M}| \hat{H}e^{-i\hat{H}t} |\F_{\rm x}\ket
\nn\\
&=\big(
\e^{c}_{k-q_{1}\ldots-q_{M}}-\e^{v}_{k}+\sum_{j=1}^{M}\w_{q_{j}}\big)Y_{kq_{1}\ldots q_{M}}(t)
\nn\\
&-\frac{1}{\callN}\sum_{p}U_{p}Y_{k+pq_{1}\ldots q_{M}}(t)
\nn\\
&+\frac{1}{\sqrt{\callN}}\sum_{j=1}^{M}\l^{k}_{q_{j}}
Y_{kq_{1}\ldots\stackrel{\sqcap}{q}_{j}\ldots q_{M}}(t)
\nn\\
&+\frac{1}{\sqrt{\callN}}
\sum_{q}\l^{k}_{q}Y_{kqq_{1}\ldots q_{M}}(t).
\label{hierarchy}
\end{align}
Already at this stage we can discuss the conditions for the 
development of an incoherent regime. Taking into account 
Eqs.~(\ref{tdPsi}) and (\ref{expansion}) we find for the electronic 
density matrix
\begin{align}
\hat{\r}_{\rm el}(t)&=
|\a|^{2}|\F_{0}\ket\bra\F_{0}|
\nn\\&+|\b|^{2}
\sum_{M=0}^{\iif}\frac{1}{M!}\sum_{q_{1}\ldots q_{M}}
|\F_{{\rm x},q_{1}\ldots q_{M}}(t)\ket\bra \F_{{\rm x},q_{1}\ldots 
q_{M}}(t)|
\nn\\
&+\left(\b\a^{\ast}\sum_{k}Y_{k}(t)|k\ket\bra\F_{0}|
+{\rm h.c.}\right),
\label{asymrhoel}
\end{align}
where we have defined 
\be
|\F_{{\rm x},q_{1}\ldots q_{M}}(t)\ket\equiv\sum_{k}Y_{kq_{1}\ldots q_{M}}(t)|k\ket.
\ee
A direct comparison with Eq.~(\ref{incoherentreg}) shows that 
for the system to relax toward an incoherent steady-state  
the zero-phonon amplitude $Y_{k}(t)$ must vanish as 
$t\to\iif$ and 
\be
\lim_{t\to\iif}\sum_{M=0}^{\iif}\frac{1}{M!}\sum_{q_{1}\ldots q_{M}}
Y_{kq_{1}\ldots q_{M}}(t)Y^{\ast}_{k'q_{1}\ldots q_{M}}(t)=
\sum_{i}w_{i}\callY^{i}_{k}\callY^{i\ast}_{k'},
\label{cond2phoel}
\ee
where $\callY_{k}^{i}$ are some $k$-dependent and time-independent complex 
quantities. We shall comment on the fulfillment of these properties 
using the exact solution.

The hierarchy in Eq.~(\ref{hierarchy}) can be solved analytically in some special, yet 
relevant, cases. The first condition to meet is 
\be
\e^{c}_{k-q_{1}\ldots-q_{M}}=\e^{c}_{k}
\label{cond1}
\ee
for all $\{q_{1},\ldots,q_{M}\}$ and for all $M$. Rigorously this 
condition is satisfied only for a perfectly flat conduction band. 
However, Eq.~(\ref{cond1}) is a good approximation 
for couplings $\l$'s and frequencies 
$\w$'s such that the distribution of phonon-momenta has a peak around 
$q=0$ with standard deviation $\s_{Q}$  much 
smaller than the momentum-scale 
over which the dispersion $\e^{c}_{k}$ varies. No restrictions on the 
dispersion of the valence band $\e^{v}_{k}$ are imposed.

Let us introduce the vectors
\be
(\blY_{q_{1}\ldots q_{M}})_{k}=Y_{kq_{1}\ldots q_{M}}
\ee
and the matrices
\be
\mathbb{H}_{kk'}=\d_{kk'}(\e^{c}_{k}-\e^{v}_{k})-\frac{U_{p-k}}{\callN}
\quad;\quad
\mathbb{L}_{q,kk'}=\d_{kk'}\l^{k}_{q}.
\ee
Then the hierarchy in Eq.~(\ref{hierarchy}) 
can be 
rewritten in matrix form as (omitting the 
dependence on time)
\begin{align}
i\dot{\blY}_{q_{1}\ldots q_{M}}&=
(\mathbb{H}+\w_{q_{1}}\ldots+\w_{q_{M}})\blY_{q_{1}\ldots q_{M}}
\nn\\
&+\sum_{j=1}^{M}\frac{\mathbb{L}_{q_{j}}}{\sqrt{\callN}}
\blY_{q_{1}\ldots\stackrel{\sqcap}{q}_{j}\ldots q_{M}}
+\sum_{q}\frac{\mathbb{L}_{q}}{\sqrt{\callN}}\blY_{qq_{1}\ldots q_{M}}.
\label{hierarchy2}
\end{align}
Notice that the matrix $\mathbb{H}$ is the Bethe-Salpeter Hamiltonian 
for an $eh$ pair. Hence the spectrum of $\mathbb{H}$ consists of a 
discrete excitonic part with subgap eigenvalues and a continuum $eh$ part
with eigenvalues larger than the gap.

\subsection{$q$-independent coupling}
\label{qsubsec}

We consider 
optical phonons $\w_{q}=\w_{0}$ and coupling matrices 
$\mathbb{L}_{q}=\mathbb{L}$ depending only on the electron momentum, hence 
$\l^{k}_{q}=\l^{k}$. In this case all coefficients of the hierarchy 
are independent of the phonon momenta. We look for solutions of the 
form
\be
\blY_{q_{1}\ldots q_{M}}(t)=\sqrt{\frac{M!}{\callN^{M}}}
\blY_{M}(t).
\label{optsol}
\ee
Inserting Eq.~(\ref{optsol}) into Eq.~(\ref{hierarchy2}) we  
find a closed system of equations for the $\blY_{M}$'s
\be
i\dot{\blY}_{M}=(\mathbb{H}+M\w_{0})\blY_{M}+\sqrt{M}\mathbb{L}
\blY_{M-1}+\sqrt{M+1}\mathbb{L}\blY_{M+1}.
\label{opticalhierarchy}
\ee
This simplified hierarchy can easily be solved with continued matrix 
fraction techniques~\cite{Martinez_2003,Stefanuccipumping}. 
The time evolved state in  Eq.~(\ref{expansion}) 
then reads
\be
|\F_{\rm 
x}(t)\ket=\sum_{M=0}^{\iif}\sum_{k}\frac{Y_{kM}(t)}{\sqrt{M!}}|kM\ket,
\ee
where
\be
|kM\ket\equiv \frac{1}{\sqrt{\callN^{M}}}\sum_{q_{1}\ldots 
q_{M}}|kq_{1}\ldots q_{M}\ket.
\ee
For arbitrary initial conditions $\blY_{M}(0)=\d_{M0}\blY^{0}$ 
no steady state is ever attained in the long time limit.
 This means that 
the electronic density matrix 
does not evolve toward an admixture like in Eq.~(\ref{incoherentreg}).
Nonetheless, the spectral function $A_{k}(T,\w)$ becomes independent 
of $T$ as $T\to\iif$ if the probing time, i.e., the $\t$-window of 
integration in Eq.~(\ref{spectralfunction}), is much longer 
than $2\p/\w_{0}$.

It is instructive to expand the vectors $\blY_{M}$ in eigenvectors 
$\blY^{(\m)}$ with energies $E^{(\m)}$ of the Bethe-Salpeter Hamiltonian 
$\mathbb{H}$
\be
\blY_{M}(t)=\sum_{\m}\a_{\m M}(t)\blY^{(\m)}.
\label{expansionamuM}
\ee
The hierarchy in Eq.~(\ref{opticalhierarchy}) can be used to 
obtain a hierarchy for the coefficients of the expansion
\begin{align}
i\dot{\a}_{\m M}&=(E^{(\m)}
+M\w_{0})\a_{\m M}\nn\\& +\sum_{\n}L_{\m\n}
\left[\sqrt{M}\a_{\n M-1}+\sqrt{M+1}\a_{\n M+1}\right],
\label{hierarchyalpha}
\end{align}
where we have defined 
\be
L_{\m\n}\equiv \blY^{(\m)\dag}\mathbb{L}\blY^{(\n)}.
\label{Lbsebasis}
\ee
If the coupling $\l^{k}$ depends on $k$ then 
the matrix element $L_{\m\n}$ is, in general, nonvanishing for $\m\neq \n$. 
This implies that even if there are no phonons  at time $t=0$  and the 
zero-phonon component 
$Y_{0}(0)=Y^{(\m_{0})}$ is an eigenstate of 
$\mathbb{H}$, hence $\a_{\m M}(0)=\d_{0M}\d_{\m_{0}\m}$,  
electronic states  with $\n\neq \m_{0}$ are visited during the 
evolution.

\subsection{$k$-independent coupling}

The hierarchy in Eq.~(\ref{hierarchy2}) can be solved exactly 
also in the special case $\mathbb{L}_{q}= \l_{q}\mathbb{1}$,
i.e., 
for an electron-phonon coupling $\l^{k}_{q}=\l_{q}$ independent 
of the electronic momentum. No restrictions on the phonon frequencies is necessary in 
this case. We define
\be
\blY_{q_{1}\ldots q_{M}}(t)=e^{-i\mathbb{H}t}\blX_{q_{1}\ldots 
q_{M}}(t),
\label{YvsX}
\ee
and rewrite the hierarchy for the vectors 
$\blX$'s
\begin{align}
i\dot{\blX}_{q_{1}\ldots q_{M}}&=
(\w_{q_{1}}\ldots+\w_{q_{M}})\blX_{q_{1}\ldots q_{M}}
\nn\\
&+\sum_{j=1}^{M}\frac{\l_{q_{j}}}{\sqrt{\callN}}
\blX_{q_{1}\ldots\stackrel{\sqcap}{q}_{j}\ldots q_{M}}
+
\sum_{q}\frac{\l_{q}}{\sqrt{\callN}}\blX_{qq_{1}\ldots q_{M}}.
\label{hierarchy2X}
\end{align}
We look for solutions of the form
\be
\blX_{q_{1}\ldots q_{M}}(t)=f_{q_{1}}(t)\ldots f_{q_{M}}(t)\blX(t).
\label{Xsol}
\ee
Inserting Eq.~(\ref{Xsol}) into Eq.~(\ref{hierarchy2X}) we find that 
the hierarchy is solved provided that 
\be
i\dot{f}_{q}=\frac{\l_{q}}{\sqrt{\callN}}+\w_{q}f_{q},
\label{langreth}
\ee
and
\be
i\dot{\blX}=\sum_{q}\frac{\l_{q}}{\sqrt{\callN}}f_{q}(t)\blX.
\label{dotx}
\ee
The solution of Eq.~(\ref{langreth}) with boundary condition $f_{q}(0)=0$ (no 
phonons at the initial time) is the Langreth function~\cite{Langreth.1970}
\be
f_{q}(t)=\frac{1}{\sqrt{\callN}}\frac{\l_{q}}{\w_{q}}(e^{-i\w_{q}t}-1).
\label{fq(t)}
\ee
Substituting the explicit form of $f_{q}(t)$ into Eq.~(\ref{dotx}) we 
find
\be
\blX(t)=\ell(t)\blX(0),
\ee
with
\be
\ell(t)=\exp\left[\frac{1}{\callN}\sum_{q}\left(\frac{\l_{q}}{\w_{q}}\right)^{2}
(-1+e^{-i\w_{q}t}+i\w_{q}t)\right].
\ee
Taking into account Eqs.~(\ref{YvsX}) and (\ref{Xsol}) we eventually 
get
\be
\blY_{q_{1}\ldots q_{M}}(t)=
f_{q_{1}}(t)\ldots f_{q_{M}}(t)\ell(t)\blY^{0}(t),
\label{case2sol}
\ee
where $\blY^{0}(t)=e^{-i\mathbb{H}t}\blY^{0}$ and 
$(\blY^{0})_{k}=Y^{0}_{k}$ are the amplitudes of the state 
$|\F_{\rm x}\ket$. 
In  Eq.~(\ref{case2sol}) the phonon dynamics is 
decoupled from the  electron dynamics. If the initial state is an 
eigenstate of $\mathbb{H}$, i.e., $\blY^{0}=\blY^{(\m_{0})}$ then no 
other electronic state is visited at later times.

For acoustic phonons the function $\ell(t)$ vanishes as $t\to\iif$ and 
therefore $Y_{k}(t)=\ell(t)Y^{0}_{k}(t)$ vanishes too in the same 
limit. This implies that the electronic density matrix becomes an 
admixture in the long-time limit, see Eq.~(\ref{asymrhoel}). To 
determine the nature of this admixture we have to evaluate 
Eq.~(\ref{cond2phoel}). Taking into account that 
$|\ell(t)|^{2}=\exp[-\sum_{q}|f_{q}(t)|^{2}]$ it is straightforward 
to find
\be
\lim_{t\to\iif}\hat{\r}_{\rm el}(t)=
|\a|^{2}|\F_{0}\ket\bra\F_{0}|
+|\b|^{2}|\F_{\rm x}^{(0)}(t)\ket\bra\F_{\rm x}^{(0)}(t)|
\ee
with $|\F_{\rm x}^{(0)}(t)\ket=\sum_{k}Y^{0}_{k}(t)|k\ket$. Therefore 
the admixture contains only two states and for it to attain a steady 
state the initially pumped state must be an eigenstate of the 
electron Hamiltonian $\mathbb{H}$. In particular if the initial state 
is the excitonic eigenstate then $Y^{0}_{k}(t)=e^{-iE^{(\rm 
x)}t}Y_{k}^{(\rm x)}$ and 
\be
\lim_{t\to\iif}\hat{\r}_{\rm el}(t)=
|\a|^{2}|\F_{0}\ket\bra\F_{0}|
+|\b|^{2}|\F_{\rm x}\ket\bra\F_{\rm x}|.
\ee
Comparing this result with the electronic density matrix at the 
initial time, see Eqs.~(\ref{virtex}) and (\ref{purestate}), we 
conclude that 
the two-band model is able to describe the complete loss of 
coherence due to scattering of electrons with acoustic phonons.

\section{Resonant pumping}
\label{resonantsec}

\begin{figure*}[tbp]
\includegraphics[width=0.95\textwidth]{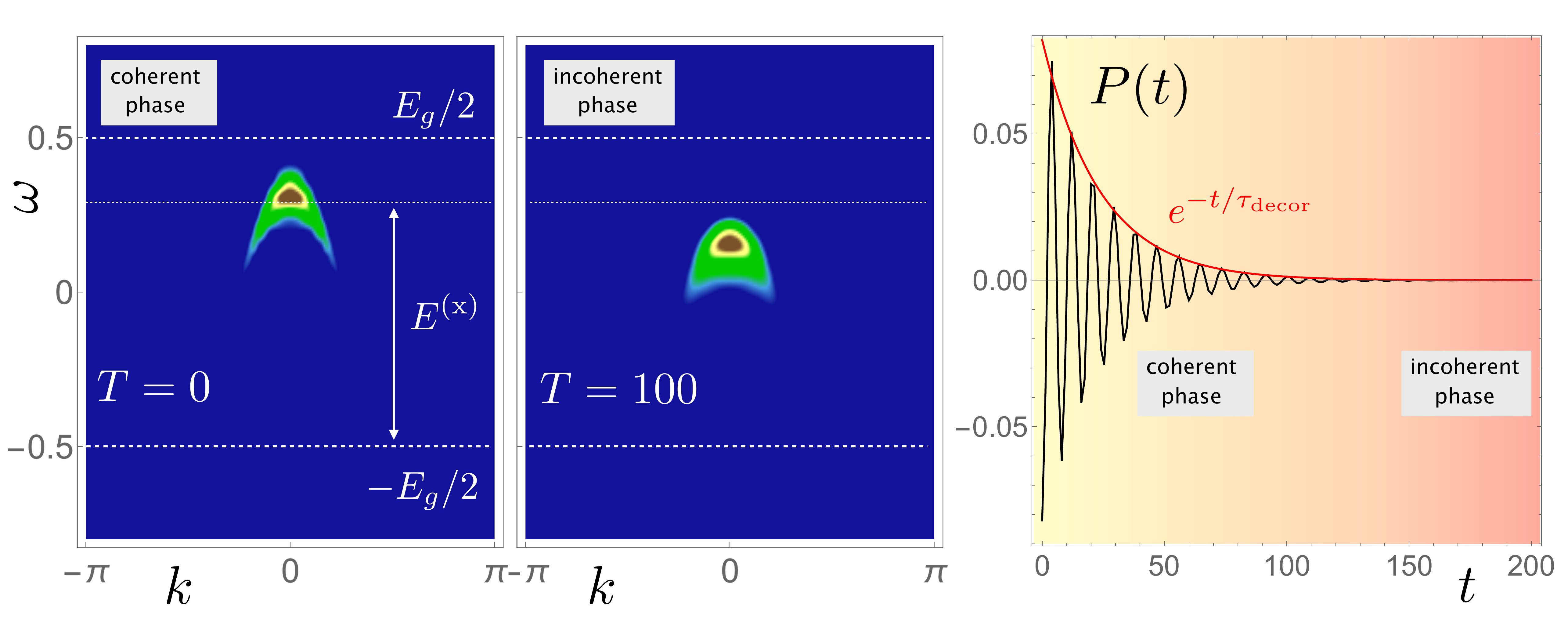}
\caption{Spectral function of the interacting electron-phonon system 
for $T=0$ (left) and $T=100$ (middle). Tick-dashed lines have been 
drawn in correspondence of the CBM and VBM. Thin-dashed lines have 
been drawn in correspondence of the exciton energy. The right panel 
shows the time-dependent polarization calculated with $d_{k}=d$ for 
all $k$-points.}
\label{acustic-spectrum}
\end{figure*} 

We investigate the effects of phonon dressing and decoherence in
a semiconductor driven by coherent light of 
frequency in resonance with the energy of a bright exciton (resonant pumping).
For illustration we consider a 1D model with a flat conduction band 
$\e_{k}^{c}=\e_{c}=E_{g}/2$ [conduction band minimum (CBM) at 
$E_{g}/2$] and a dispersive valence band 
$\e^{v}_{k}=E_{g}[\cos k-1]-E_{g}/2$ [valence band maximum (VBM) at 
$-E_{g}/2$] separated by an energy gap 
$E_{g}$. Henceforth all energies are measured in units of $E_{g}$ 
and times in units of $1/E_{g}$. 
For a short-range (momentum-independent) Coulomb interaction $U_{p}=U=0.65$ the electronic 
system admits an exciton state $\blY^{(\rm x)}$ of energy $E^{(\rm 
x)}\simeq 0.81$.
Under the condition of resonant pumping the initial state is therefore 
$\blY^{0}=\blY^{(\rm x)}$.

Without electron-phonon scattering, i.e., $\l^{k}_{q}=0$, 
the spectral function $A_{k}(T,\w)$ in Eq.~(\ref{spectralfunction})
is independent of 
$T$ and for each $k$ it exhibits a single peak at energy 
$\e^{v}_{k}+E^{(\rm 
x)}$~\cite{Kremp_PhysRevB.78.125315,PSMS.2016,RustagiKemper2018,PSMS.2019}, see also 
Appendix~\ref{Gccapp}:
\be
A_{k}(T,\w)=2\p |\b|^{2}\left|Y^{(\rm x)}_{k}\right|^{2}\d(\w-\e^{v}_{k}-E^{(\rm 
x)}).
\label{A0x}
\ee
This spectral function can be interpreted as the 
fully interacting spectral function just after the excitation, 
i.e., at $T=0$ 
-- phonon scatterings kick in only at later times.
Equation~(\ref{A0x})
yields an excitonic sideband;
it is a replica of the valence band located at energy $E^{(\rm x)}$ above the 
VBM with spectral 
weight proportional to the square of the excitonic wavefunction.
In Fig.~\ref{acustic-spectrum} (left) we show the color plot of 
Eq.~(\ref{spectralfunction}). Calculations have been performed with $\callN=81$ $k$-points using a $\t$-window 
of integration that extends from $\t_{\rm min}=-60$ to $\t_{\rm 
max}=-\t_{\rm min}$. 

The electron-phonon coupling causes the spectral 
function to acquire a dependence on  $T$, even for 
initial states that are eigenstates of $\mathbb{H}$.
We study the evolution of the system when 
conduction electrons interact through a coupling 
$\l_{q}=\frac{\l_{0}}{2}(\cos q+1)$ with acoustic phonons of dispersion 
$\w_{q}=\w_{s}\sin|q|$. In the numerical simulations we 
have chosen $\l_{0}=0.18$ and $\w_{s}=2(E_{g}-E^{(\rm x)})=0.38$ -- which 
is twice the exciton binding energy. In Appendix~\ref{Gcvapp} we 
demonstrate that the polarization $P(t)$ in Eq.~(\ref{polariz}) can 
be written as
\be
P(t)=\a^{\ast}\b\frac{1}{\callN}\sum_{k}d_{k}Y_{k}(t)+{\rm h.c.}
\label{polacustic}
\ee
where $Y_{k}(t)=\ell(t)e^{-iE^{(\rm x)}t}Y^{0}_{k}$, see Eq.~(\ref{case2sol}).
Let us briefly discuss the behavior of $P(t)$ for a vanishing 
electron-phonon coupling. In this case $\ell(t)=1$ and the 
polarization  oscillates monochromatically 
at frequency $E^{(\rm x)}$. 
In the  diagrammatic nonequilibrium Green's function theory a simple 
Hartree-Fock (HF) treatment of the Coulomb 
interaction is enough to reproduce
the oscillatory behavior~\cite{Schmitt-Rink_PhysRevB.37.941,Ostreich_1993,Schafer_ZPB1986,Lindberg_PhysRevB.38.3342,PSMS.2019}.
Interestingly, the HF theory also predicts the appearance of the excitonic 
sideband in the spectral function, see Eq.~(\ref{A0x}). 
The coherent oscillations of $P(t)$ are crucial to observe 
this effect, which does indeed originate from 
the eigenvalues of 
the HF Floquet Hamiltonian~\cite{Perfetto_PhysRevLett.125.106401}. 
Switching on the electron-phonon coupling 
the function $\ell(t)$, and hence 
the amplitude  $Y_{k}(t)$, vanishes for $t\to\iif$.
In Fig.~\ref{acustic-spectrum} (right) we show $P(t)$ for a system 
with dipole matrix elements 
$d_{k}=d$ for all $k$. 
For $t\simeq 100$ the polarization is suppressed by about two order 
of magnitudes.
We shall show that in this incoherent phase (times $t>100$)
the exact spectral function still exhibits  
an excitonic sideband. 
Therefore excitonic structures in the spectral function are
not a hallmark of excitonic coherence,
as it is erroneously predicted by the HF theory.

To obtain the spectral 
function  in the incoherent regime we 
evaluated Eq.~(\ref{spectralfunction}), see Appendix~\ref{Gccapp}, at $T=100$ using the same $\t$-window 
of integration as in the left panel of Fig.~\ref{acustic-spectrum}. The 
result is shown in the middle panel of the same figure. 
The excitonic sideband is still clearly visible although 
it is slightly shifted downward. In fact, it 
would be more appropriate to refer to this spectral feature as the 
{\em exciton-polaron sideband}. The experiment 
of Ref.~\cite{man2020experimental} has most likely measured such 
sideband as the system was probed after 0.5 picoseconds from the 
resonant excitation.
We also observe that phonon-dressing is 
responsible for an energy broadening of the sideband, consistently with 
the reported damping of the polarization.

The energy shift of the excitonic sideband as the system evolves from the 
coherent to the incoherent phase can also be related to the  Stokes 
shift~\cite{toyozawa_2003}.
At zero momentum
the coherent peak is detected at energy $E^{(\rm 
x)}$, i.e., at the onset of the photoabsorption spectrum, whereas the 
incoherent peak is red shifted by the ``reorganization energy'', 
i.e., the energy gain in the transition from an exciton to an
exciton-polaron. Therefore the energy of the incoherent peak lies at the 
onset of the photoluminescence spectrum~\cite{Li.jpclett.8b03604,Hoyer_PhysRevB.72.075324}.
This also agrees with the fact 
that the photoluminescence signal is proportional to the  
population of (incoherent) excitons~\cite{Hannewald_PhysRevB.62.4519}. 
In this respect trARPES provides a unique investigation tool 
as it allows for extracting information which would otherwise require 
two independent experiments, i.e., absorption and luminescence.

\begin{figure}[tbp]
\includegraphics[width=0.4\textwidth]{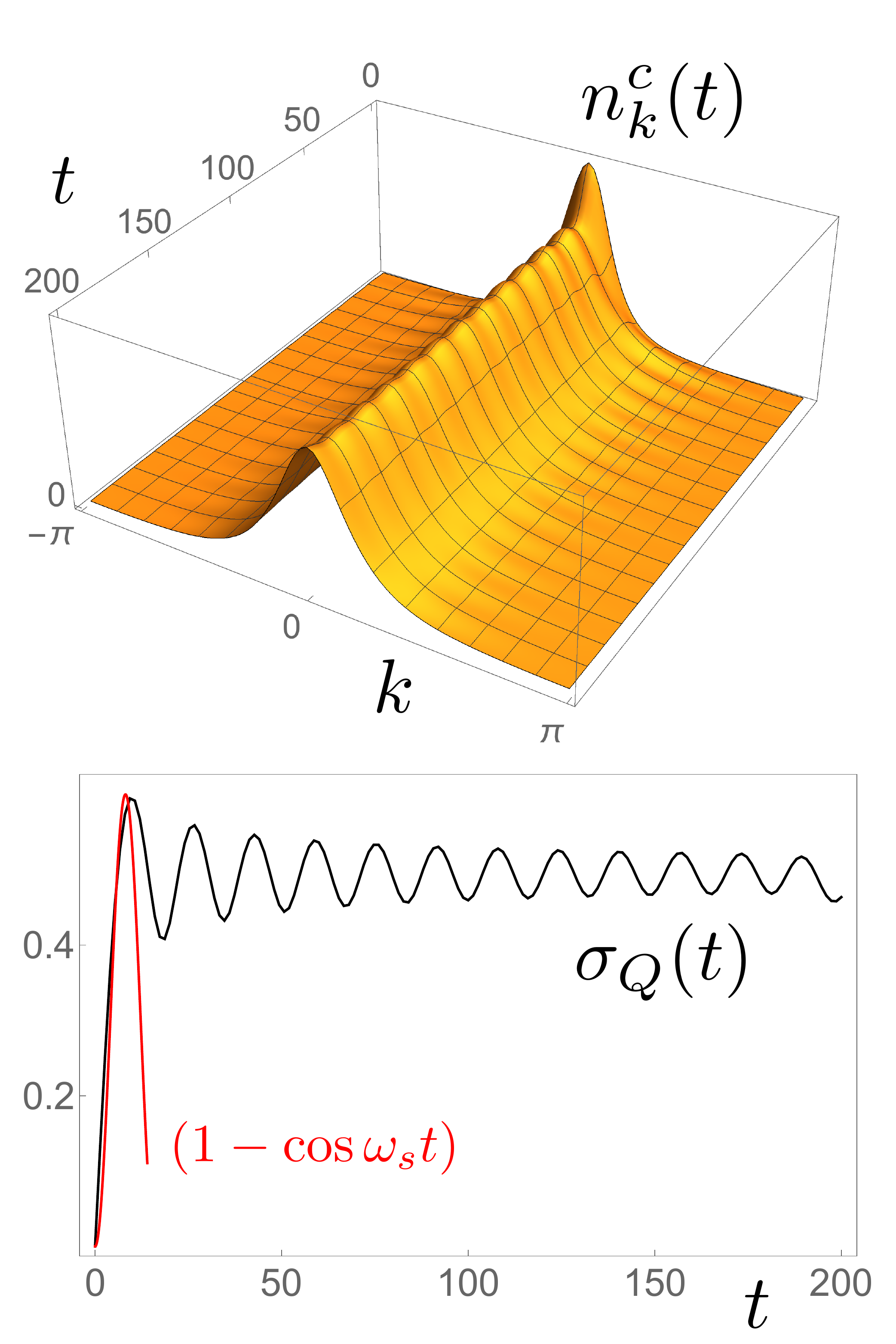}
\caption{Top: Time-dependent $k$-resolved electronic occupations in the 
conduction band. Bottom: Time-dependent standard deviation of the 
phonon momentum. }
\label{exciton-polaron-acustic}
\end{figure} 

In Fig.~\ref{exciton-polaron-acustic} (top) we show the evolution of the 
$k$-resolved electron density in the conduction band. At time $t=0$ the profile 
is proportional to the square of the exciton wavefunction in momentum 
space, i.e., $n^{c}_{k}(0)\propto |Y^{(\rm x)}_{k}|^{2}$, see 
Appendix~\ref{Gccapp}. As time 
increases the exciton transforms into an exciton-polaron and the 
wavefunction slightly spreads. Thus the dressed $eh$ wavepacket becomes 
more localized in real-space, in agreement with the larger binding 
energy observed in  Fig.~\ref{acustic-spectrum} (middle).
We observe that the phonon dressing 
occurs on a much faster time-scale than the decoherence time, 
i.e., the timescale over which the polarization damps. This can also 
be seen from the plot of the time-dependent standard deviation 
$\s_{Q}(t)$, see Fig.~\ref{exciton-polaron-acustic} (bottom). 
According to Eq.~(\ref{sigmaQacustic}) the standard deviation grows 
like $\s_{Q}(t)\propto (1-\cos \w_{s}t)$. Hence the 
timescale $\t_{\rm dress}$ for the phonon dressing is
\be
\t_{\rm dress}=\frac{2\p}{\w_{s}}\simeq 16.3.
\ee
The envelope function  of the polarization can be found from 
Eq.~(\ref{polacustic}). After some simple algebra one finds an 
approximate exponential decay $\exp[-t/\t_{\rm decoh}]$
with decoherence time 
\be
\t_{\rm decoh}=\frac{\w_{s}}{2\l_{0}^{2}}=23.8.
\ee
In Fig.~\ref{acustic-spectrum} (right) the envelope function nicely 
interpolates all maxima of the time-dependent polarization, see red 
curve. 
We emphasize that the dressing time is dictated by the largest phonon 
frequency whereas 
the decoherence time is dictated by the  smallest
polaronic shift.  
Furthermore,  the dressing timescale does not appear in an exponential 
function, see again Fig.~\ref{exciton-polaron-acustic}.
We also notice that $\s_{Q}(t)\simeq   
0.5$ in the long-time limit;
hence, similar results would have been obtained for a dispersive 
conduction band with a shallow enough minimum.

Although the electronic occupations as well as the polarization 
attain a steady value in the long time limit the phononic occupations 
do not. This  behavior is a direct consequence of 
the acoustic dispersion and the low dimensionality. 
In Appendix~\ref{nphapp} we demonstrate that $n_{q}(t)=|f_{q}(t)|^{2}$
where the function  $f_{q}(t)$ is given in Eq.~(\ref{fq(t)}).
Therefore  the total number of phonons at 
time $t$ is given by $N_{\rm 
ph}(t)=\frac{2}{\callN}\sum_{q}(\frac{\l_{q}}{\w_{q}})^{2}(1-\cos\w_{q}t)$. 
For large times the main contribution to the sum comes from low-momentum phonons.
Approximating $\w_{q}\simeq \w_{s}|q|$, $\l_{q}\simeq \l_{0}$ and taking the
thermodynamic limit $\callN\to\iif$ we then find
\be
N_{\rm ph}(t)\simeq 
\frac{\l_{0}^{2}}{\w_{s}}\,t=\frac{1}{2}\frac{t}{\t_{\rm decoh}}.
\ee
Thus the total number of phonons increases linearly in time. It is 
easy to show that the divergence of $N_{\rm ph}(t)$ at $t\to\iif$ is 
milder in two dimensions, being it $\log(t)$, and it is absent in 
three dimensions. We also observe that the free-phonon contribution 
to the total energy $E_{\rm ph}(t)=\sum_{q}\w_{q}n_{q}(t)$ diverges 
like $\log(t)$ in one dimension whereas it approaches a constant 
value for larger dimensions.

\section{Nonresonant pumping}
\label{nonresonantsec}

Electrons pumped in the conduction band have enough energy to emit 
optical phonons and hence to decay into a bound exciton state. 
The issue we intend to address here is whether the spectral function 
exhibits an excitonic structure inside the gap and, in the 
affirmative case, what the shape is.

As already pointed out the exact solution of the two-band model Hamiltonian does not attain a steady state for 
optical phonons, see discussion in Section~\ref{qsubsec}. However, probing 
the system with pulses of duration 
much longer than the inverse of the phonon frequency the spectral 
function $A_{k}(T,\w)$ becomes independent of $T$ as $T\to\iif$.
We consider again a 1D model 
with a flat conduction band and a dispersive valence band separated 
by an energy gap $E_{g}$: 
$\e_{k}^{c}=\e_{c}=E_{g}/2$ (CBM at $E_{g}/2$) and
$\e^{v}_{k}=E_{g}[\cos k-1]-E_{g}/2$ (VBM at $-E_{g}/2$).
We also consider the same 
short-range Coulomb interaction $U_{p}=0.65$ (all energies are in 
units of $E_{g}$) as in the previous section -- hence the system 
admits one exciton state at energy $E^{(\rm x)}=0.31$. Let us assume 
that the pump has excited the system in a wavepacket of 
continuum $eh$ states $\blY^{(c)}$ of the Bethe-Salpeter 
Hamiltonian $\mathbb{H}$ with energy $E^{(c)}\simeq E_{g}$, hence
\be
\blY^{0}=\frac{1}{\sqrt{\callN_{0}}}\sum_{c:E^{(c)}\simeq E_{g}}\blY^{(c)},
\label{y0init}
\ee
where $\callN_{0}$ is the number of eigenstates in the sum. In the 
thermodynamic limit the ratio $r_{0}=\callN_{0}/\callN\ll 1$ remains 
finite.
At zero pump-probe delay, i.e., $T=0$, we can ignore 
phonon effects and the spectral function reads [compare with the resonant case 
Eq.~(\ref{A0x})]:
\be
A_{k}(0,\w)=2\p|\b|^{2}
\Big|Y^{0}_{k}\Big|^{2}\d(\w-E_{g}/2).
\ee
As  $Y^{0}_{k}$ is peaked around $k=0$ the spectral function 
$A_{k\simeq 0}(0,\w)$ is peaked at frequency $\w\simeq 
E_{g}/2=\e^{c}$ (CBM), and it is vanishingly small for nonvanishing 
momenta.

\begin{figure}[tbp]
\includegraphics[width=0.45\textwidth]{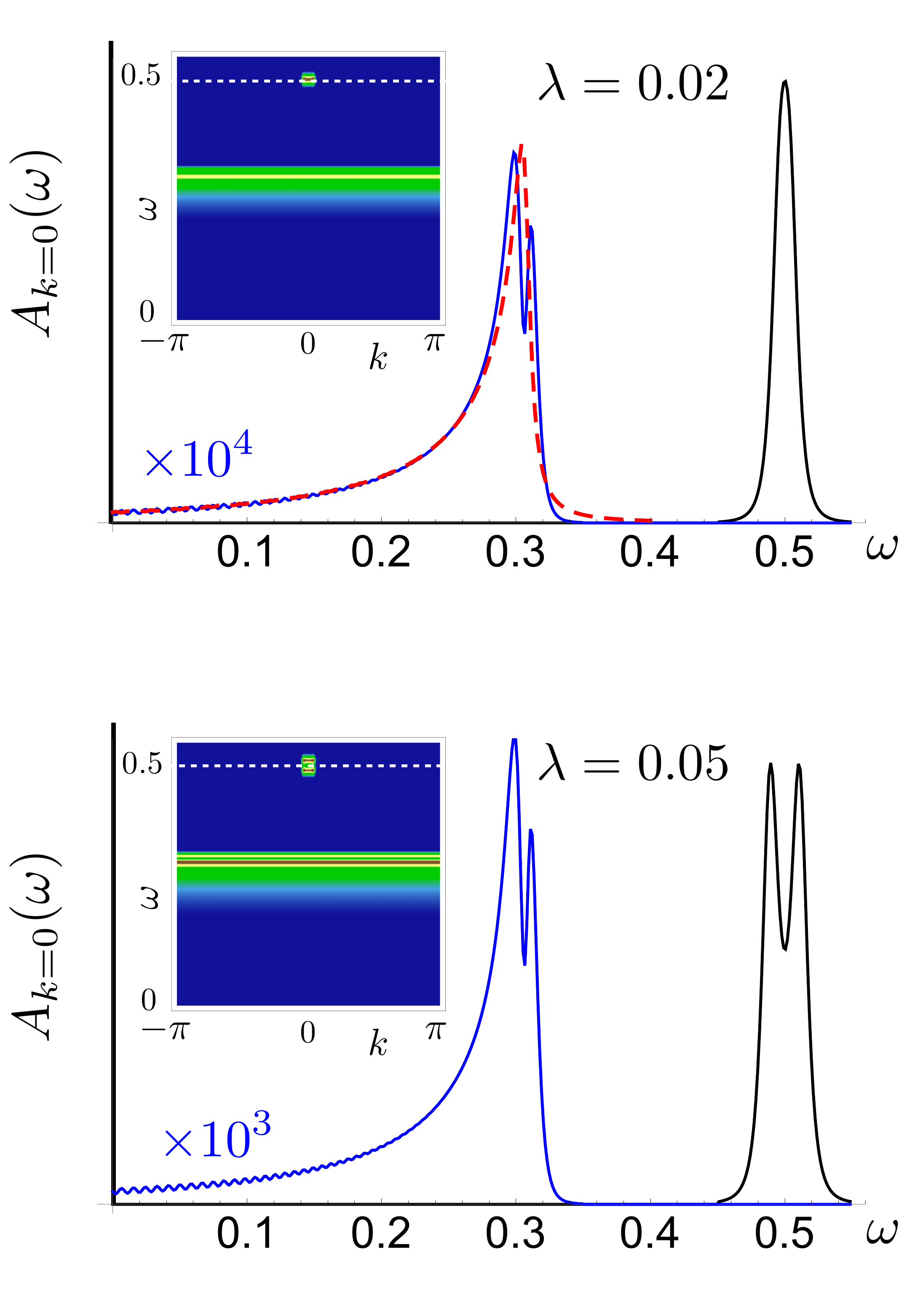}
\caption{Spectral function $A_{k\simeq 0}(\w)$ in the long-time limit 
$T\to\iif$ for a weak ($\l_{0}=0.02$, top panel) and intermediate 
($\l_{0}=0.05$, bottom panel) electron-phonon coupling. The exciton 
contribution (blue line) is magnified by a factor $10^{4}$ (top) and 
$10^{3}$ (bottom). The dashed (red) line in the top panel is the 
exciton wavefunction in energy space, see main text.
The insets show the color plot of the spectral 
function in the $k-\w$ plane.}
\label{exciton-polaron-optical}
\end{figure} 

In Appendix~\ref{Gccapp} we show that for electrons coupled to a 
branch of optical phonons of frequency $\w_{0}$
the spectral function in the long-time limit
becomes
\begin{align}
\lim_{T\to\iif}A_{k}(T,\w)&=\lim_{T\to\iif}\frac{|\b|^{2}}{T}
\times
\Big\{\Big|\tilde{Y}_{k0}(\w-\e^{v}_{k})\Big|^{2}
\nn\\
&+\sum_{M=1}^{\iif}\frac{1}{\callN}\sum_{p}
\Big|\tilde{Y}_{pM}(\w-\e^{v}_{p}+M\w_{0})\Big|^{2}
\Big\}.
\label{spectralfunction1}
\end{align}
The amplitudes $\tilde{\blY}_{M}(\w)$ are the Fourier transform of 
the time-dependent amplitudes $\blY_{M}(t)$ 
in Eq.~(\ref{opticalhierarchy}).
Interestingly only the first term ($M=0$) depends on $k$. All 
remaining terms ($M\geq 1$)
contribute with a $k$-independent function of the frequency 
(a consequence of the nondispersive nature of the conduction band).
Expanding the amplitudes like in 
Eq.~(\ref{expansionamuM}) we can equivalently calculate 
$\tilde{\blY}_{M}(\w)$ from the Fourier transform 
of the hierarchy in Eq.~(\ref{hierarchyalpha}).
We neglect here the couplings $L_{cc'}$ between low-energy continuum 
states since these scatterings are suppressed by energy conservation.
We instead consider the couplings $L_{c\rm x}=L_{{\rm 
x}c}=\l_{0}/\sqrt{\callN}$ between the exciton state and the 
continuum states. We then find
\begin{align}
i\dot{\a}_{c M}&=(E^{(c)}
+M\w_{0})\a_{c M}\nn\\& +\frac{\l_{0}}{\sqrt{\callN}}
\left[\sqrt{M}\a_{{\rm x} M-1}+\sqrt{M+1}\a_{{\rm x} M+1}\right],
\label{hierarchyalphac}
\end{align}
\begin{align}
i\dot{\a}_{{\rm x} M}&=(E^{(\rm x)}
+M\w_{0})\a_{{\rm x} M}\nn\\& +\frac{\l_{0}}{\sqrt{\callN}}
\sum_{c}\left[\sqrt{M}\a_{c M-1}+\sqrt{M+1}\a_{c M+1}\right],
\label{hierarchyalphax}
\end{align}
to be solved with boundary conditions 
$\a_{cM}(0)=\d_{M0}/\sqrt{\callN_{0}}$ if $E^{(c)}\simeq E_{g}$ and 
zero otherwise, see Eq.~(\ref{y0init}).
The symmetry of the electron-phonon coupling preserves the symmetry 
of the initial state, i.e., $\a_{cM}(t)$ is independent of $c$ if $E^{(c)}\simeq 
E_{g}$ and $\a_{cM}(t)=0$ otherwise. We can therefore simplify the 
hierarchy by introducing the quantity $\a_{0M}(t)=\sum_{c}\a_{cM}(t)$ 
and by taking into account that $E^{(c)}\simeq E_{g}$ for all $c$ in 
the initial wavepacket.
To gain some insight  we solve the simplified  hierarchy to lowest order in $\l_{0}$, 
i.e., we neglect processes with the emission of two or more phonons. 
These processes are relevant for strong electron-phonon coupling 
and give rise to phononic replica 
bands~\cite{Moser_PhysRevLett.110.196403,Story_PhysRevB.90.195135,chen2015observation,Antonius_PhysRevB.92.085137,wang2016tailoring,verdi2017origin,Hubner_2018}.
Fourier transforming Eqs.~(\ref{hierarchyalphac}) and 
(\ref{hierarchyalphax}) we find
\begin{align}
\tilde{\a}_{c0}(\w)&=-\frac{1}{\sqrt{\callN_{0}}}\,
\Im\left[
\frac{1}{\w-E_{g}-\frac{r_{0}\l_{0}^{2}}{\w-E^{(\rm x)}-\w_{0}}}
\right],
\\
\tilde{\a}_{{\rm x}1}(\w)&=-\l_{0}r_{0}\,
\Im\left[
\frac{1}{(\w-E_{g}-r_{0}\l_{0}^{2})(\w-E^{(\rm x)}-\w_{0})}
\right],
\label{ax1}
\end{align}
and $\tilde{\a}_{c1}(\w)=\tilde{\a}_{{\rm x}0}=0$. Using these 
solutions the asymptotic spectral function of 
Eq.~(\ref{spectralfunction1}) becomes
\begin{align}
\lim_{T\to\iif}A_{k}(T,\w)&= \lim_{T\to\iif}\frac{|\b|^{2}}{T}
\Big\{\Big|\tilde{\a}_{c0}(\w+E_{g}/2)Y^{0}_{k}\Big|^{2}
\nn\\
&+\frac{1}{\callN}\sum_{p}\Big|
\tilde{\a}_{{\rm x}1}(\w-\e^{v}_{p}+\w_{0})Y^{(\rm x)}_{p}\Big|^{2}
\Big\}
\label{workedoutA}
\end{align}

In Fig.~\ref{exciton-polaron-optical} we show the spectral function for
a system with $\callN=501$ $k$-points, $r_{0}=0.05$ 
and phonon frequency $\w_{0}=E_{g}-E^{\rm x}\simeq 0.2$. 
In the weak coupling regime $\l_{0}=0.02$ (top panel)
the quasi-particle peak at the CBM (frequency $\w=E_{g}/2$, black curve) is 
accompanied by an excitonic structure (blue curve) having asymmetric lineshape and 
offset at frequency $\w\simeq E^{(\rm x)}$. The main effect of an 
increased electron-phonon coupling is the splitting of the quasi-particle peak 
(energy splitting $\simeq \l_{0}^{2}r_{0}$) and a larger intensity of 
the excitonic structure, see bottom panel. 

In both (weak and 
intermediate) regimes 
the momentum and energy dispersion of the excitonic structure is 
considerably different from the resonant case, compare insets of 
Fig.~\ref{exciton-polaron-optical} with  
left and middle panels of Fig.~\ref{acustic-spectrum}. 
We point out that this marked qualitative difference is not related to the 
phonon dispersion. For resonant pumping a replica of the valence band would have 
emerged even with optical phonons. 
The results in 
Fig.~\ref{exciton-polaron-optical} clearly indicate that 
incoherent excitons forming after nonresonant pumping 
give rise to a replica of the {\em conduction} band, thus confirming 
the nonequilibrium $T$-matrix Green's function 
treatment~\cite{Asano-2014,PSMS.2016,Steinhoff2017}.

It is also worth commenting on the asymmetric lineshape  at 
fixed $k$ (blue line) of the excitonic structure. 
This feature too was observed in the nonequilibrium $T$-matrix Green's function 
treatment~\cite{PSMS.2016,Steinhoff2017}. 
The exact solution allows for a precise characterization of it. 
The excitonic sideband originates from the second term in Eq.~(\ref{workedoutA}). Taking 
into account the explicit form of $\tilde{\a}_{{\rm x}1}(\w)$ in 
Eq.~(\ref{ax1}) we see that the asymmetric lineshape is well 
described by  
$\callL(\w)\equiv \sum_{p}\d(\w-E^{(\rm x)}-\e^{v}_{p})|Y^{(\rm x)}_{p}|^{2}$. 
The function $\callL$ is shown in the top panel of 
Fig.~\ref{exciton-polaron-optical} (dashed line) and it can be 
interpreted 
as the excitonic wavefunction in ``energy space''.

\section{Summary}
\label{consec}

We have calculated the trARPES spectra of a  two-band model 
semiconductor from the exact analytic solution of the time-dependent  many-body 
wavefunction. Both electron-electron and electron-phonon 
interactions have been taken into account, and the exact solution has been 
worked out for acoustic as well as optical phonons. 
Numerical results have been presented to address open issues on 
the spectra of resonantly and nonresonantly excited semiconductors after 
phonon-induced decoherence and relaxation.

In resonantly excited semiconductor the initial excitonic sideband 
is a replica of the valence band located at the onset of the 
photoabsorption spectrum (with respect to the VBM). After phonon-induced decoherence 
the excitonic sideband changes slightly but it does not fade away.
In particular its position is red-shifted (Stokes shift) ending up at the 
onset of the photoluminescence spectrum (with respect to the VBM).
Phonon-dressing is also responsible for a broadening of 
the  sideband, in agreement with the transition from 
excitons to exciton-polarons. We also find that phonon dressing 
occurs on a much faster time-scale than phonon-induced decoherence, 
the former being dictated by the largest phonon frequency whereas the 
latter by the smallest polaronic shift. In the incoherent phase the 
electronic subsystem is in a stationary state, i.e., the electronic 
occupations and the electronic Green's 
function are invariant under time translations. The stationarity of 
the phononic subsystem does instead depend on the dimensionality; 
the number of low-momentum acoustic phonons grows in time linearly  in 
one-dimension, logarithmically in two dimensions and attains a 
constant value in larger dimensions.

In nonresonantly excited semiconductor the excitonic sideband forms 
only after phonon-driven relaxation. Its shape is a replica of the 
conduction band and  its energy distribution at fixed momentum  
has a highly asymmetric lineshape. The lineshape 
is proportional to the exciton wavefunction in energy space.

The considered two-band model ignores several aspects of real materials, 
e.g., multiple bands and valleys, intraband and interband long-range 
Coulomb interactions, band anisotropies and degeneracies, multiple phonon 
branches, etc.~\cite{Sangalli2019}. However, the addressed issues are independent of these 
details. In Ref.~\cite{PSMS.2019} we have shown that 
the excitonic features of the ARPES spectrum of a bulk LiF in the resonant phase 
could be predicted using the same model. The scenario
emerging from our exact solution is therefore expected to be 
generally applicable to the interpretation of experimental trARPES spectra.

\begin{acknowledgments}
We  acknowledge the financial support from MIUR PRIN (Grant 
No. 20173B72NB), from INFN through the TIME2QUEST project, and from 
Tor Vergata University through the Beyond Borders Project 
ULEXIEX.      
\end{acknowledgments}

\appendix

\section{Conduction-conduction Green's function}
\label{Gccapp}

Let us begin with the calculation of the conduction-conduction  
Green's function defined in Eq.~(\ref{Gcc}). 
For convenience we rewrite it  as 
\be
G^{cc}_{k}(t,t')=i|\b|^{2}\r^{cc}_{k}(t,t'),
\ee
where $\r^{cc}_{k}(t,t')\equiv\bra\F_{\rm x}(t')|
\hat{c}^{\dag}_{k}
e^{-i\hat{H}(t'-t)}\hat{c}_{k}|\F_{\rm x}(t)\ket$.
Using the expansion in Eq.~(\ref{expansion}) we have
\begin{align}
\r_{k}^{cc}(t,t')&=\sum_{pp'}\sum_{M=0}^{\iif}\frac{1}{(M!)^{2}}
\sum_{\substack{q_{1}\ldots q_{M}\\q'_{1}\ldots q'_{M}}}
Y^{\ast}_{p'q'_{1}\ldots q'_{M}}(t')Y_{pq_{1}\ldots q_{M}}(t)\quad
\nn\\
&\times
\bra p'q'_{1}\ldots 
q'_{M}|\hat{c}^{\dag}_{k}
e^{-i\hat{H}(t'-t)}\hat{c}_{k}|pq_{1}\ldots q_{M}\ket.
\label{rhok1}
\end{align}
For vanishing electron-phonon coupling $Y_{k}(t)=Y^{0}_{k}(t)$ and 
$Y_{pq_{1}\ldots q_{M}}(t)=0$ for all $M\geq 1$. We thus recover the purely 
electronic result
\be
\r^{0cc}_{k}(t,t')\equiv\lim_{\{\l^{k}_{q}\}\to 0}\r_{k}^{cc}(t,t')
=e^{i\e^{v}_{k}(t'-t)}Y^{0\ast}_{k}(t')Y^{0}_{k}(t).
\label{erhocc}
\ee
If the initial state $\blY^{0}$ 
is an eigenstate $\blY^{(\m)}$ of $\mathbb{H}$ with energy $E^{(\m)}$ 
then $Y^{0}_{k}(t)=Y^{(\m)}_{k}e^{-iE^{(\m)}t}$ and Eq.~(\ref{erhocc}) 
becomes
\be
\r^{0cc}_{k}(t,t')=|Y^{(\m)}_{k}|^{2}e^{-i(\e^{v}_{k}+E^{(\m)})(t-t')}.
\label{r0cck}
\ee
In this case the spectral function $A_{k}(T,\w)$ in Eq.~(\ref{spectralfunction})
is independent of 
$T$ and for each $k$ it exhibits a single peak at energy 
$\e^{v}_{k}+\e^{(\m)}$:
\be
A_{k}(T,\w)=2\p |\b|^{2}|Y^{(\m)}_{k}|^{2}\d(\w-\e^{v}_{k}-E^{(\m)}).
\label{A0}
\ee

To proceed with the calculation of the Green's function of the 
coupled electron-phonon system we observe that
the state $\hat{c}_{k}|pq_{1}\ldots q_{M}\ket$ in Eq.~(\ref{rhok1}) is a 
state with no electrons in the conduction band and with a hole of 
momentum $p$ in the valence band. Therefore
\be
e^{i\hat{H}\t}\hat{c}_{k}|pq_{1}\ldots 
q_{M}\ket=e^{i(-\e^{v}_{p}+\w_{q_{1}}\ldots+\w_{q_{M}})\t}
\hat{c}_{k}|pq_{1}\ldots 
q_{M}\ket.
\label{pq1qMt}
\ee
Furthermore
\begin{align}
\bra p'q'_{1}\ldots q'_{M}|\hat{c}^{\dag}_{k}\hat{c}_{k}|pq_{1}\ldots q_{M}\ket
&=\d_{p'p}\sum_{P}\prod_{j=1}^{M}\d_{q'_{j}q_{P(j)}}
\nn\\
&\times \d_{k,p-q_{1}\ldots-q_{M}}.
\label{c+kck}
\end{align}

\subsection{$q$-independent coupling}

We insert Eqs.~(\ref{pq1qMt}) and (\ref{c+kck}) into 
Eq.~(\ref{rhok1}) and use the solution in Eq.~(\ref{optsol}) for the 
amplitudes. We find
\begin{align}
\r_{k}^{cc}(t,t')&=\sum_{q_{1}\ldots q_{M}}
\d_{k,p-q_{1}\ldots-q_{M}}
\nn\\
&\times \!\!\sum_{p}\sum_{M=0}^{\iif}\frac{1}{\callN^{M}}Y^{\ast}_{pM}(t')Y_{pM}(t)
e^{i(\e^{v}_{p}-M\r_{0})(t'-t)}
\nn\\
&=Y^{\ast}_{k0}(t')Y_{k0}(t)e^{i\e^{v}_{k}(t'-t)}
\nn\\&+
\frac{1}{\callN}\sum_{p}\sum_{M=1}^{\iif}Y^{\ast}_{pM}(t')Y_{pM}(t)
e^{i(\e^{v}_{p}-M\r_{0})(t'-t)}.
\label{rhokopt}
\end{align}
For vanishing electron-phonon coupling only the first term on the 
right hand side contributes and we recover Eq.~(\ref{erhocc}).

Let us study the steady-state limit of Eq.~(\ref{rhokopt}). We define 
the center-of-mass time $T=(t+t')/2$ and the relative time 
$\t=(t-t')$. Given a function $a(t)$ with Fourier transform 
$\tilde{a}(\w)$ we have
\begin{align}
\lim_{T\to\iif}a(t)a^{\ast}(t')&=
\int \frac{d\w}{2\p}e^{-i\w\t}
\nn\\ &\times \lim_{T\to\iif}
\int\frac{d\W}{2\p}e^{-i\W 
T}\tilde{a}(\w+\frac{\W}{2})\tilde{a}^{\ast}(\w-\frac{\W}{2})
\nn\\
&=\lim_{T\to\iif}\frac{1}{T}\int \frac{d\w}{2\p}
e^{-i\w\t}|\tilde{a}(\w)|^{2}
\end{align}
where in the last equality we used the Riemann-Lebesgue theorem. 
Using this result in Eq.~(\ref{rhokopt}) with functions 
$a(t)=Y_{pM}(t)e^{-i(\e^{v}_{p}-M\r_{0})t}$
and taking into account Eq.~(\ref{spectralfunction}) for the spectral 
function we find Eq.~(\ref{spectralfunction1}).

\subsection{$k$-independent coupling}

Proceeding along the same lines as for the $q$-independent coupling 
but using the solution in
Eq.~(\ref{case2sol}) we find
\begin{align}
\r_{k}^{cc}(t,t')&=\sum_{p}e^{i\e^{v}_{p}(t'-t)}Y^{0\ast}_{p}(t')Y^{0}_{p}(t)
\ell^{\ast}(t')\ell(t)
\sum_{M}\frac{1}{M!}
\nn\\
&\times \!\!\sum_{q_{1}\ldots q_{M}}
\d_{k,p-q_{1}\ldots-q_{M}}\prod_{j=1}^{M}f^{\ast}_{q_{j}}(t')f_{q_{j}}(t)
e^{-i\w_{q_{j}}(t'-t)}.
\label{rhok2}
\end{align}
We define the function
\be
h_{q}(t,t')\equiv f^{\ast}_{q}(t')f_{q}(t)
e^{-i\w_{q}(t'-t)},
\ee
and its Fourier expansion
\be
h_{q}(t,t')=\sum_{n}e^{iqn}\,\frac{\tilde{h}_{n}(t,t')}{\callN}.
\label{hqft}
\ee
Depending on the dimensionality $D$ of the system the product 
$qn\equiv\sum_{i=1}^{D}q_{i}n_{i}$ stands for the scalar product between 
the vector $q$ in the first Brillouin zone and the position $n$ of 
the unit cell. The domain over which $n$ runs  is such that 
$\frac{1}{\callN}\sum_{n}e^{i(q-q')n}=\d_{qq'}$ and 
$\frac{1}{\callN}\sum_{q}e^{iq(n-n')}=\d_{nn'}$. We can then use the 
identity 
\be
\sum_{q_{1}\ldots q_{M}}\!\!\d_{k,p-q_{1}\ldots-q_{M}}
\prod_{j=1}^{M}h_{q_{j}}(t,t')=\sum_{n}e^{i(p-k)n}
\frac{[\tilde{h}_{n}(t,t')]^{M}}{\callN}
\ee
to perform the sum over $M$ in Eq.~(\ref{rhok2}), obtaining 
the following compact expression
\be
\r_{k}^{cc}(t,t')=\sum_{p}K_{k-p}(t,t') 
\,e^{i\e^{v}_{p}(t'-t)}Y^{0\ast}_{p}(t')Y^{0}_{p}(t).
\label{rho3}
\ee
In Eq.~(\ref{rho3}) the kernel
\be
K_{k-p}(t,t')\equiv \ell^{\ast}(t')\ell(t)\frac{1}{\callN}\sum_{n}e^{i(p-k)n}
e^{\tilde{h}_{n}(t,t')}
\label{kernel}
\ee
is the only quantity  depending on 
the electron-phonon couplings and phonon 
frequencies. 

For vanishing electron-phonon coupling, i.e., 
$\l_{q}=0$ for all $q$, 
the Langreth function $\ell(t)=1$, see Eq.~(\ref{langreth}), and the 
function $\tilde{h}_{n}(t,t')=0$. Therefore 
\be
K_{k-p}^{0}(t,t')\equiv \lim_{\{\l_{q}\}\to 0}K_{k-p}(t,t')=\d_{kp}
\ee
and Eq.~(\ref{rho3}) correctly reduces to the purely electronic 
result in Eq.~(\ref{erhocc}).
The conduction-conduction  Green's function of the interacting 
electron-phonon system is a convolution in momentum space between the Green's function of 
the purely electronic system $\r^{0cc}$ and the kernel $K$. Indeed  
Eq.~(\ref{rho3}) can be rewritten as
\be
\r^{cc}_{k}(t,t')=\sum_{p}K_{k-p}(t,t')\r^{0cc}_{p}(t,t').
\ee

For the system to attain a steady state in the long-time limit the 
kernel $K_{k-p}(t,t')$ has to approach a function of $\t=t-t'$ 
for $t,t'\to\iif$.
If we denote by $K_{k-p}(\w)$ its Fourier transform 
then the steady-state spectral 
function in Eq.~(\ref{spectralfunction}) reads
\be
\lim_{T\to\iif}A_{k}(T,\w)=
\sum_{p}\int\frac{d\w'}{2\p}K_{k-p}(\w-\w')A^{0}_{p}(\w')
\label{ssaacustic}
\ee
where $A^{0}_{p}(\w')$ is the spectral function of the purely 
electronic system.

\section{Conduction-valence Green's function}
\label{Gcvapp}

The calculation of the off-diagonal Green's function in 
Eq.~(\ref{Gcv}) is straightforward. We start by rewriting it as
\be
G^{cv}_{k}(t,t')=i\a^{\ast}\b\r^{cv}_{k}(t,t'),
\ee
where $\r^{cv}_{k}(t,t')\equiv \bra\F_{0}|
\hat{v}^{\dag}_{k}
e^{-i\hat{H}(t'-t)}\hat{c}_{k}|\F_{\rm x}(t)\ket$.
Inserting the expansion in Eq.~(\ref{expansion}) and using 
Eq.~(\ref{pq1qMt}) we find 
\begin{align}
\r^{cv}_{k}(t,t')&=
\sum_{p}\sum_{M=0}^{\iif}\frac{1}{M!}
\sum_{q_{1}\ldots q_{M}}Y_{pq_{1}\ldots q_{M}}(t)
\nn\\
&\times e^{i(\e^{v}_{p}-\w_{q_{1}}\ldots-\w_{q_{M}})(t'-t)}
\bra\F_{0}|\hat{v}^{\dag}_{k}\hat{c}_{k}
|pq_{1}\ldots q_{M}\ket.
\end{align}
The bracket in this equation is nonvanishing only for $M=0$ and 
$k=p$, in which case its value is unity. Hence
\be
\r^{cv}_{k}(t,t')=Y_{k}(t)e^{i\e^{v}_{k}(t'-t)}.
\label{rhocvgen}
\ee
For $k$-independent coupling this result further simplifies since 
$Y_{k}(t)=\ell(t)Y^{0}_{k}(t)$, see Eq.~(\ref{case2sol}).

\section{Time-dependent phonon occupations and momentum distribution}
\label{nphapp}

The time-dependent phonon occupancy is defined in Eq.~(\ref{nq(t)}).
Taking into account the expansion in Eq.~(\ref{expansion}) 
we have
\begin{align}
n_{q}(t)&=\sum_{M=0}^{\iif}\frac{1}{(M!)^{2}}\sum_{kk'}
\sum_{\substack{q_{1}\ldots q_{M}\\q'_{1}\ldots q'_{M}}}
Y^{\ast}_{k'q'_{1}\ldots q'_{M}}(t)Y_{kq_{1}\ldots q_{M}}(t)
\nn\\
&\times
\bra k'q'_{1}\ldots 
q'_{M}|\hat{b}_{q}^{\dag}\hat{b}_{q}|kq_{1}\ldots q_{M}\ket.
\end{align}
From the inner product in Eq.~(\ref{innerproduct}) it follows that
\begin{align}
\bra k'q'_{1}\ldots 
q'_{M}|\hat{b}_{q}^{\dag}\hat{b}_{q}|kq_{1}\ldots q_{M}\ket&=
\d_{k'k}\sum_{P}\prod_{j=1}^{M}\d_{q'_{j}q_{P(j)}}
\nn\\
&\times\sum_{j=1}^{M}\d_{qq_{j}}.
\end{align}
Using the total symmetry of the amplitudes we then get
\begin{align}
n_{q}(t)&=\sum_{M=0}^{\iif}\frac{1}{M!}
\sum_{k,q_{1}\ldots q_{M}}\sum_{j=1}^{M}\d_{qq_{j}}
|Y_{kq_{1}\ldots q_{M}}(t)|^{2}.
\label{nqgeneral}
\end{align}

We are also interested in the phonon-momentum distribution. 
The phonon-momentum operator is given by
$\hat{Q}=\sum_{q}q\,\hat{n}_{q}$.
The average momentum of the exciton-polaron state is therefore
\begin{align}
Q(t)&=\bra\F_{\rm x}(t)|\hat{Q}|\F_{\rm x}(t)\ket
\nn\\
&=\sum_{M=0}^{\iif}\frac{1}{M!}
\sum_{k,q_{1}\ldots q_{M}}\sum_{j=1}^{M}q_{j}
|Y_{kq_{1}\ldots q_{M}}(t)|^{2},
\end{align}
whereas the standard deviation is
\begin{align}
\s^{2}_{Q}(t)&=\bra\F_{\rm x}(t)|\hat{Q}^{2}|\F_{\rm x}(t)\ket-Q^{2}(t)
\nn\\
&= \sum_{M=0}^{\iif}\frac{1}{M!}
\sum_{k,q_{1}\ldots q_{M}}\sum_{ij=1}^{M}q_{i}q_{j}
|Y_{kq_{1}\ldots q_{M}}(t)|^{2}-Q^{2}(t).
\end{align}
The condition in Eq.~(\ref{cond1}) is satisfied provided that 
$Q(t)\simeq 0$ and 
$\s_{Q}(t)$ is much smaller than the momentum-scale over which the conduction 
band changes.

\subsection{$q$-independent coupling}
Let us evaluate Eq.~(\ref{nqgeneral}) using the solution of Eq.~(\ref{optsol}) for  
$q$-independent couplings. It is straightforward to find
\begin{align}
n_{q}(t)&=\sum_{M=0}^{\iif}M\times
\sum_{k}\frac{|Y_{kM}(t)|^{2}}{\callN}.
\label{nqcase1}
\end{align}
Hence the phonon occupancy depends on 
the initial electronic state $\blY^{0}$ but 
it does not  depend of $q$ (all modes are equally 
populated). This implies that $Q(t)=0$ and hence the standard 
deviation is simply
\be
\s^{2}_{Q}(t)=\sum_{q}q^{2}\sum_{M=0}^{\iif}M\times
\sum_{k}\frac{|Y_{kM}(t)|^{2}}{\callN}.
\ee

\subsection{$k$-independent coupling}

For $k$-independent couplings we substitute  the  solution of 
Eq.~(\ref{case2sol}) and find
\begin{align}
n_{q}(t)&=\sum_{M=0}^{\iif}\frac{1}{M!}
M|f_{q}(t)|^{2}\left(\sum_{q'}|f_{q'}(t)|^{2}\right)^{M-1}|\ell(t)|^{2}
\nn\\
&=|f_{q}(t)|^{2},
\label{nq(t)final}
\end{align}
where we have taken into account that $\sum_{k}|Y^{0}_{k}(t)|^{2}
=\blY^{0\dag}e^{i\mathbb{H}t}e^{-i\mathbb{H}t}\blY^{0}=1$.
In this case the phonon occupancy in Eq.~(\ref{nq(t)final}) 
is {\em independent} of the initial state.

Assuming that $f_{q}(t)=f^{\ast}_{-q}(t)$ the average momentum 
$Q(t)=0$ for all times. Therefore the standard deviation reads
\be
\s^{2}_{Q}(t)=\sum_{q}q^{2}|f_{q}(t)|^{2}.
\label{sigmaQacustic}
\ee


\end{document}